\documentclass[aps,pra,amsmath,showpacs,twocolumn]{revtex4}

\usepackage{hyperref}
\usepackage{graphicx}
\usepackage{natbib}

\graphicspath{{figs/}}

\newcommand{\func}[2]{\ensuremath{#1\left(#2\right)}}
\newcommand{\Thpp}{\ensuremath{\mathrm{Th}^{2+}}}
\newcommand{\Thppp}{\ensuremath{\mathrm{Th}^{3+}}}
\newcommand{\ddt}{\ensuremath{\frac{d}{dt}}}

\begin{document}

\title{Charge Exchange and Chemical Reactions with Trapped Th$^{3+}$}
\date{\today}
\author{L.~R.~Churchill} \altaffiliation[Permanent address: ]{Applied
  Physics Laboratory, Johns Hopkins University, Laurel, Maryland
  20723}
\author{M.~V.~DePalatis}
\author{M.~S.~Chapman}
\affiliation{School of Physics, Georgia Institute of Technology,
  Atlanta, Georgia 30332-0430}


\begin{abstract}
  We have measured the reaction rates of trapped, buffer gas cooled
  $\Thppp$ and various gases and have analyzed the reaction products
  using trapped ion mass spectrometry techniques. Ion trap lifetimes
  are usually limited by reactions with background molecules, and the
  high electron affinity of multiply charged ions such as $\Thppp$
  make them more prone to loss. Our results show that reactions of
  $\Thppp$ with carbon dioxide, methane, and oxygen all occur near the
  classical Langevin rate, while reaction rates with argon, hydrogen,
  and nitrogen are orders of magnitude lower. Reactions of $\Thppp$
  with oxygen and methane proceed primarily via charge exchange, while
  simultaneous charge exchange and chemical reaction occurs between
  $\Thppp$ and carbon dioxide. Loss rates of $\Thppp$ in helium are
  consistent with reaction with impurities in the gas. Reaction rates
  of $\Thppp$ with nitrogen and argon depend on the internal
  electronic configuration of the $\Thppp$.
\end{abstract}

\maketitle

\section{Introduction}
\label{sec:introduction}

Unlike most atomic nuclei, which have excitation energies in the range
of keV to MeV, the nucleus of the thorium isotope, $^{229}$Th, has an
excited state just several eV above the nuclear ground state
\cite{Beck2007}.  The transition between the nuclear ground state and
this unique isomeric state lies within the UV optical spectrum, where
it can be addressed using coherent light sources.  The coherent
control of the electronic states of atoms with tunable lasers has been
a major focal point of modern atomic physics.  Extending this paradigm
to the control of nuclear states of atoms would represent a
significant achievement.  The transition between the $^{229}$Th
nuclear states could potentially be used as a frequency reference with
a fractional uncertainty approaching $10^{-20}$ \cite{Peik2003}.
Furthermore, the transition may be especially sensitive to changes in
the value of the fine structure constant, allowing up to 5--6 orders
of magnitude enhancement in measurements of its time variation
\cite{Flambaum2006}.  However, this latter point requires further
study \cite{Hayes2007,He2007,He2008,Berengut2009}.

\begin{figure}
  \centering
  \includegraphics{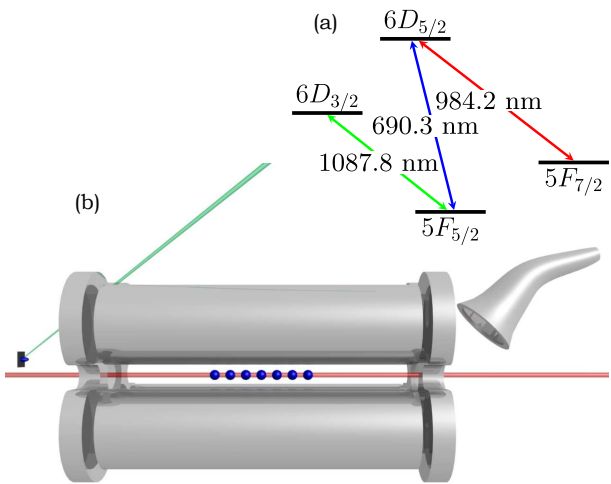}
  \caption{(Color online) (a) $\Thppp$ energy levels and optical
    transitions. (b) Schematic drawing of ion trap system.  A target
    outside of the ion trap (left) is ablated using the third harmonic
    of a pulsed Nd:YAG laser (diagonal line).  Lasers used for optical
    excitation of the trapped ions are aligned along the trap axis
    (horizontal line).  A CEM located outside of the trap (right) is
    used for electronic detection.  The CEM is mounted off-axis to
    allow clear optical access.}
  \label{fig:trap-schematic}
\end{figure}

In order to demonstrate coherent control of the nuclear state, thorium
atoms must be confined in such a way that they can be continuously
interrogated and observed.  Unlike the lower ionization states, triply
ionized thorium has a convenient level structure for fluorescence
detection and laser cooling \cite{Peik2003}.  The low-lying electronic
states of $\Thppp$ and the optical transitions between them are shown
in Fig.~\ref{fig:trap-schematic}.  Using the more common thorium
isotope, $^{232}$Th, our group has demonstrated the creation,
trapping, and laser cooling of $\Thppp$ \cite{Campbell2009}.

In our earliest observations of $^{232}\Thppp$, the ions would remain
in the trap for only a few seconds.  In general, elastic collisions
with background molecules are not likely to result in ejection of ions
due to the large depth of an ion trap.  The primary modes of $\Thppp$
loss are charge exchange and chemical reactions with background gases.
By improving the background vacuum and increasing the purity of the
buffer gas used for initial cooling, we were able to extend the trap
lifetime of $\Thppp$ to minutes.  By removing the buffer gas
immediately after the initial trap loading and laser cooling the ions,
a lifetime of $>10$ minutes was obtained \cite{Campbell2009}.

The limited $\Thppp$ trap lifetime presents a significant challenge in
time-intensive experiments like measuring the hyperfine states of
$^{229}\Thppp$ and searching for the nuclear isomer transition.
Furthermore, the short lifetime represents a tremendous cost in
performing experiments with $^{229}\Thppp$ given the extraordinary
price of the isotope ($>$ \$100k/mg).

To better understand the charge exchange and chemical reaction
processes and quantify their rates, we conducted a series of
experiments to determine the reaction rate coefficients between
$\Thppp$ and various gases.  A summary of the experimental results can
be found in Table~\ref{tab:rates}.  The remainder of this paper is
devoted to describing these experiments in greater detail.

\begin{table}
  \centering
  \begin{ruledtabular}
    \begin{tabular}{c|c|cc|c}
      {\bf Reactant} & $k$ & $k_{\varepsilon}$ & ($\varepsilon$) &
      $k_L \times 10^{-9}$ \\ \hline
      He & $<3.4 \times 10^{-15}$ & $<6.0 \times 10^{-16}$ & $(0.18)$
      & $1.6$ \\ 
      Ne & $< 4.1 \times 10^{-15}$ & $< 1.2 \times 10^{-15}$ & $(0.3)$
      & $1.0$ \\ 
      Ar$^{(*)}$ & $1.3 \times 10^{-14}$ & $1.7 \times 10^{-14}$ &
      $(1.3)$ & $1.5$ \\
      N$_2^{(*)}$ & $1.6 \times 10^{-13}$ & $1.6 \times 10^{-13}$ &
      $(1.0)$ & $1.8$ \\ 
      H$_2$ & $4.6 \times 10^{-13}$ & $2.1 \times 10^{-13}$ & $(0.46)$
      & $4.4$ \\ 
      CH$_4$ & $2.6 \times 10^{-9}$ & $3.6 \times 10^{-9}$ & $(1.4)$ &
      $2.8$ \\ 
      O$_2$ & $3.7 \times 10^{-9}$ & $3.7 \times 10^{-9}$ & $(1.0)$ &
      $1.7$ \\ 
      CO$_2$ & $2.8 \times 10^{-9}$ & $4.0 \times 10^{-9}$ & $(1.4)$ &
      $1.8$ \\
    \end{tabular}
  \end{ruledtabular}
  \caption{Charge exchange and chemical reaction rate coefficients for
    $\Thppp$.  Reaction rate coefficients are given in units
    of cm$^3$s$^{-1}$.  Here, $k_L$ is the calculated Langevin rate,
    and $k$ is the experimental reaction rate coefficient.  The
    corrected reaction rate coefficient is $k_{\varepsilon} =
    \varepsilon \cdot k$, where $\varepsilon$ is the ion gauge
    correction factor for the gas in question.  Experimental reaction
    rate coefficients for helium and neon are upper bounds.  The
    accuracy of the coefficients given here is limited by the accuracy
    of the ion gauge used to measure the pressure of the reactants.
    (*) The reaction rate for Ar and N$_2$ depends on the electronic
    configuration of Th$^{3+}$.  The experimental rate refers to
    reaction from the 5F$_{5/2}$ ground state.}
  \label{tab:rates}
\end{table}


A schematic of the ion trap used in these experiments is shown in
Fig.~\ref{fig:trap-schematic}.  The trap is outfitted with a channel
electron multiplier (CEM) for electronic detection of trapped ions.
When used in combination with the well understood mass selection
characteristics of a linear ion trap, the CEM can be used to determine
if a specific species of ion is present in the trap.

The ion trap is loaded with Th$^{3+}$ via laser ablation of a thorium
metal target with the third harmonic of a pulsed Nd:YAG laser
($\lambda = 355$ nm).  The target is located near the trap axis and
oriented perpendicular to it.  The voltage on the dc endcap nearest
the ablation target is gated with the ablation pulse, dropping to
ground when the pulse is fired, and increasing to 100 V for ion
confinement.

Helium buffer gas is present in the system throughout the loading
process.  The ablated Th$^{3+}$ ions are initially moving at speeds of
~10 km/s as they approach the ion trap.  As the hot Th$^{3+}$ ions
collide with room temperature helium atoms, they lose some of their
thermal energy.  This process serves two critical purposes.  First, it
reduces the energy of some ablated Th$^{3+}$ ions as they traverse the
length of the ion trap, thereby increasing the likelihood they will be
trapped.  Secondly, it damps the motion of trapped Th$^{3+}$ ions,
allowing them to come to thermal equilibrium at some fraction of the
trap depth.  The helium pressure is commonly held between $10^{-5}$
and $10^{-4}$ torr in these experiments.  Generally, the number of
ions loaded increases with higher buffer gas pressure, while the final
equilibrium temperature of the trapped ions decreases.

\section{Helium}
\label{sec:helium}

\begin{figure}
  \centering
  \includegraphics[angle=-90]{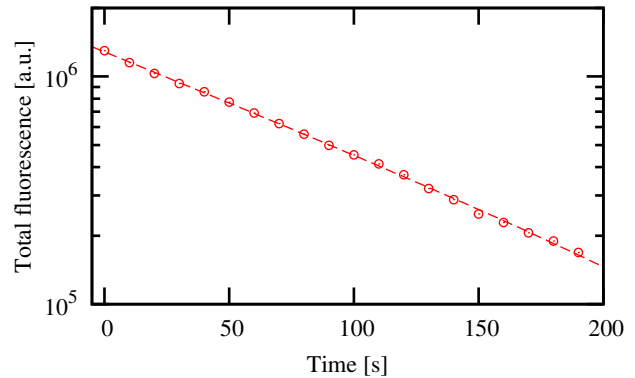}
  \caption{(Color online) Accumulated fluorescence signal for
    determination of $\Thppp$ loss rate.  This data was taken with $5
    \times 10^{-5}$ torr of helium in the vacuum chamber.  The loss
    rate was determined to be $\tau^{-1} = 0.0100(2)$ s$^{-1}$.}
  \label{fig:HeLifetime}
\end{figure}

The loss of $\Thppp$ from the ion trap in $5 \times 10^{-5}$ torr of
helium buffer gas is shown in Fig.~\ref{fig:HeLifetime}.  Here, the
$\Thppp$ decay rate is measured by observing the reduction in
fluorescence over a series of images.  This method can be used only
when the lifetime of the ions is sufficient to allow enough
fluorescence measurements for a proper exponential fit.

When the $\Thppp$ trap lifetime is too short for observation of
fluorescence, mass selective electronic detection with the CEM is
employed to measure the decay rate.  We load ions into the trap via
ablation and wait for some amount of time.  The rf and dc voltages
applied to the trap are then tuned such that only $\Thppp$ remains
stable.  The contents of the trap are then delivered to the CEM by
lowering the voltage of the trap endcap nearest it.  The time interval
before the mass selection is varied, and the data is accumulated over
several iterations for each time interval.  Many iterations are
required to suppress the variation due to fluctuations in the number
of ions loaded.  This method is typically used when the $\Thppp$
lifetime is a few seconds or less.

\begin{figure}
  \centering
  \includegraphics[angle=-90]{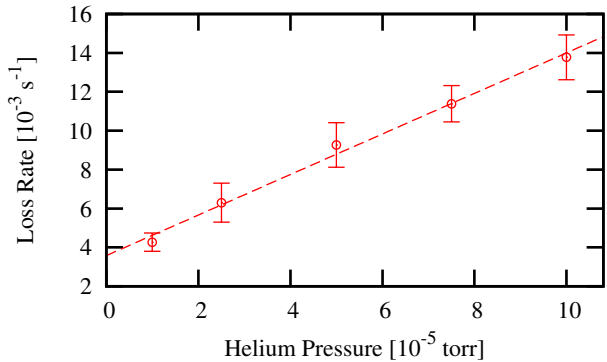}
  \caption{(Color online) $\Thppp$ loss rate as a function of BIP
    helium buffer gas pressure.  The slope of the linear fit implies a
    reaction rate coefficient of $k = 3.4 \times
    10^{-15}\,\text{cm}^{3} \text{s}^{-1}$ ($k_{\epsilon} = 6.0 \times
    10^{-16}\,\text{cm}^{3} \text{s}^{-1}$ with an ion gauge
    correction factor of $\epsilon = 0.18$), while the intercept
    implies a background loss rate of $\tau^{-1} = 3.3 \times
    10^{-3}\,\text{s}^{-1}$.  It is likely that the $\Thppp$ reacts
    with an impurity in the helium rather than with the helium itself;
    therefore, the value for $k$ sets an upper bound on the reaction
    rate coefficient for helium.}
  \label{fig:BIP-loss-rate}
\end{figure}

The decay rate through any given reaction channel is proportional to
the density of the reactant in the system. This relation can be
written as $\tau^{-1} = kn$, where $n$ is the density of the reactant,
and $k = \langle \sigma v \rangle$ is the reaction rate coefficient
for the given channel. Here, $\sigma$ is the reaction cross-section
and $v$ is the relative velocity between the reactants. For a given
reaction, the rate coefficient can be determined by measuring the
$\Thppp$ decay rate as a function of the pressure of the reactive
gas. The reaction rate coefficient is extracted from the slope of a
linear fit of this data. The intercept of the linear fit gives the
loss rate due to background gases in the vacuum. For helium, the data
and corresponding linear fit are shown in Fig.~\ref{fig:BIP-loss-rate}

Helium is a noble gas, and a charge exchange reaction between it and
$\Thppp$ is endothermic by 4.6 eV. This suggests that such a reaction
is strongly inhibited for low energy reactants. Therefore, the loss
shown in Fig.~\ref{fig:HeLifetime} is most likely due to an impurity
in the helium rather than to the helium itself.

The helium gas we use in our experiment comes from a cylinder with a
built-in purifier (BIP).  The BIP helium is supplied by Airgas, Inc.
According to the specifications, the BIP helium gas has an impurity
level $<1$ ppm.  The data of Fig.~\ref{fig:BIP-loss-rate} implies a
reaction rate coefficient between the BIP helium and Th$^{3+}$ on the
order of $10^{-16}$ to $10^{-15}$ cm$^3$s$^{-1}$.  Thus, the reaction
between Th$^{3+}$ and the contaminant within the BIP helium
responsible for its loss proceeds with a reaction rate coefficient on
the order of $10^{-10}$ to $10^{-9}$ cm$^3$s$^{-1}$.

A reaction rate coefficient of $10^{-9}$ cm$^3$s$^{-1}$ is consistent
with the value predicted by a classical collision model introduced by
Langevin \cite{Langevin1903} and refined by Gioumousis and Stevenson
\cite{Gioumousis1958}.  This model is helpful in estimating the
reaction cross-sections between ions and molecules.  We assume the ion
is a point particle with charge $Ze$, and that the molecule is
spherically symmetric with polarizability $\alpha$.  The primary
interaction between the ion and molecule is due to the electric field
of the ion and the field-induced diploe moment of the molecule.  The
interaction potential scales as $r^{-4}$.

Below threshold values of the relative velocity and the impact
parameter, the interaction potential leads to a spiraling orbit of the
ion and molecule about their shared center of mass.  If a reaction
between the ion and molecule is exothermic, it is assumed to occur
with unit probability when a spiraling orbit occurs.  The so-called
Langevin cross-section for reaction is then \cite{Gioumousis1958}
\begin{equation}
  \label{eq:Langevin-cross-section}
  \sigma_L = 2\pi \frac{Ze}{v} \sqrt{\frac{\alpha}{\mu}},
\end{equation}
where $v$ is the relative velocity between the particles and $\mu$ is
the reduced mass. The corresponding reaction rate coefficient is
\begin{equation}
  \label{eq:rxn-rate-coef}
  k_L = 2\pi Ze \sqrt{\alpha}{\mu}
\end{equation}
The preceding equations are given in Gaussian (cgs) units, which is
how they are most commonly found in the literature.

The Langevin model is not a comprehensive theoretical formulation, and
therefore, it only suffices to provide an estimate of the reaction
rate between a given ion and reactant molecule.  The model does not
apply to molecules that possess a permanent dipole moment.  In that
case, the leading term in the interaction potential scales as $r^{-2}$
rather than $r^{-4}$ \cite{Troe1985}.  Furthermore, several
experiments \cite{Schissler1956,Field1957,Boelrijk1962} have shown a
departure from the inverse velocity dependence of the classical
cross-section in some reactions.  Ultimately, the probability of a
reaction occurring in the event of a classical interaction is
dependent on the combined energy state of the reactants, the available
energy states in the final products, and the possible reaction
pathways between them \cite{Mahan1971}.  Several experiments
\cite{Johnsen1978,Johnsen1979,Johnsen1980,Holzscheiter1981,Church1989},
including our own, have found that some reaction rates can differ
significantly depending on the internal energy state of a reactant.

Nevertheless, the Langevin rate is useful in setting an upper bound on
the reaction rate between ions and non-polar molecules
\cite{McDaniel1970}.  Many ion-molecule reactions do occur at or near
the Langevin rate
\cite{Johnsen1979,Johnsen1974,Holzscheiter1981a,Chatham1983,Church1992,Andrews2005}.
The Langevin rates for the reactions we investigated are given in
Table~\ref{tab:rates} with the experimental results.

\section{Gases with low reaction rates}
\label{sec:low-rate-gases}

\begin{figure}
  \centering
  \includegraphics[angle=-90]{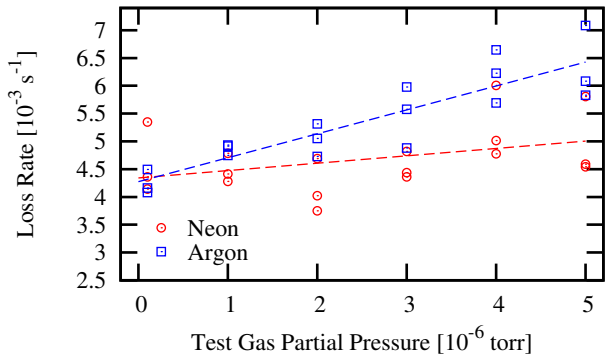}
  \caption{(Color online) $\Thppp$ loss rate in the presence of neon
    and argon.  A partial pressure of $10^{-5}\,\text{torr}$ of helium
    was added to the test gas to cool the ions for fluorescence
    observation.  The reaction rate coefficient for neon is $k = 4.1
    \times 10^{-15}\,\text{cm}^{3}\text{s}^{-1}$ ($k_{\epsilon} = 1.2
    \times 10^{-15}\,\text{cm}^{3}\text{s}^{-1}$ with an ion gauge
    correction factor of $\epsilon = 0.30$).  The decay rates shown
    for argon represent $\Thppp$ loss from the $5\text{F}_{5/2}$
    ground state.  The reaction rate coefficient is $k = 1.3 \times
    10^{-14}\,\text{cm}^{3}\text{s}^{-1}$ ($k_{\epsilon} = 1.7 \times
    10^{-14}\,\text{cm}^{3}\text{s}^{-1}$ with an ion gauge correction
    factor of $\epsilon = 1.3$).}
  \label{fig:NeonArgon}
\end{figure}

\begin{figure}
  \centering
  \includegraphics[angle=-90]{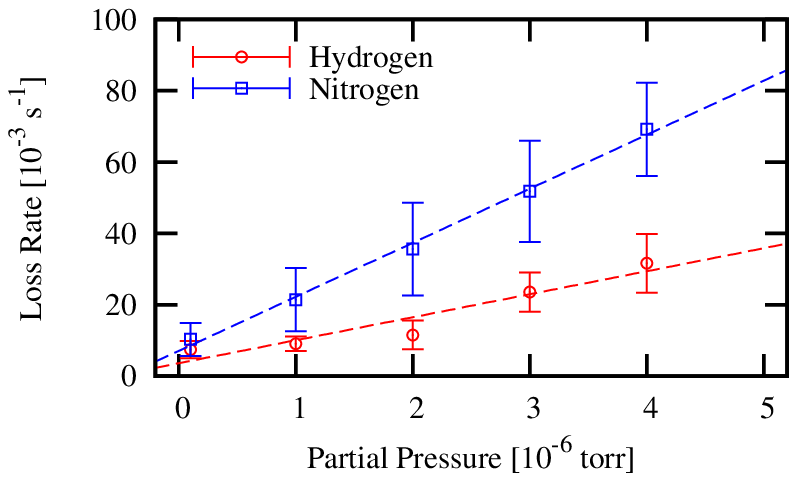}
  \caption{(Color online) $\Thppp$ loss rate in the presence of
    hydrogen and nitrogen.  A partial pressure of
    $10^{-5}\,\text{torr}$ of helium was added to the test gas to cool
    the ions for fluorescence observation. The reaction rate
    coefficient for hydrogen is $k = 4.6 \times
    10^{-13}\,\text{cm}^{3}\text{s}^{-1}$ ($k_{\epsilon} = 2.1 \times
    10^{-13}\,\text{cm}^{3}\text{s}^{-1}$ with an ion gauge correction
    factor of $\epsilon = 0.46$).  The decay rates shown for nitrogen
    represent $\Thppp$ loss from the $5\text{F}_{5/2}$ ground state.
    The reaction rate coefficient is $k =
    1.6\times10^{-13}\,\text{cm}^{3}\text{s}^{-1}$.}
  \label{fig:HNLossRates}
\end{figure}

Loss rates of Th$^{3+}$ in the presence of neon and argon are shown in
Fig.~\ref{fig:NeonArgon}, while loss rates in the presence of nitrogen
and hydrogen are shown in Fig.~\ref{fig:HNLossRates}.  For all of
these decay measurements, a partial pressure of $10^{-5}$ torr of
helium was added to the system to cool the ions for fluorescence
imaging.  As can be seen from Fig.~\ref{fig:BIP-loss-rate}, this
quantity of helium does not add significantly to the background loss
rate of Th$^{3+}$.  The reactions with argon, nitrogen, and hydrogen
all proceed relatively slowly, with reaction rate coefficient orders
of magnitude below the corresponding Langevin rates.  Since our
interest was in gases that react with Th$^{3+}$ near the Langevin
rate, we did not attempt to identify the products of these reactions.

Like helium, neon is a noble gas with a high ionization energy that
inhibits charge exchange reactions with $\Thppp$. Such reactions are
in this case endothermic by 1.6 eV. As can be seen in
Fig.~\ref{fig:NeonArgon}, $\Thppp$ loss rates in neon were never
significantly larger than what could be attributed to the presence of
the helium buffer gas.  In addition, the reaction rate coefficient we
determined from the data is consistent with reaction occurring between
the $\Thppp$ and 0.5 ppm impurity in the gas.

It is worth noting that the kinetic energy of $\Thppp$ due to rf
micromotion could conceivably contribute to overcome an otherwise
endothermic energy gap \cite{Major1968,Wu1993,DeVoe2009}, such as with
helium and neon. However, such micromotion-enhanced reactions would
require micromotion amplitudes much larger than is likely. While some
contribution of the micromotion to the loss rates is possible for
endothermic reactions, our results are consistent with quoted
impurities \cite{Airgas2009}.

Argon is also a noble gas, but a charge exchange reaction between it
and $\Thppp$ is exothermic by 4.2 eV. The $\Thppp$ reaction rates with
argon and nitrogen were found to depend on the electronic
configuration of the $\Thppp$. The data in Fig.~\ref{fig:NeonArgon}
represent loss rates from the 5F$_{5/2}$ ground state of $\Thppp$.
The 5F$_{5/2}$ state was isolated for study by shuttering the 690 nm
laser between fluorescence measurements. This allows $\Thppp$ in the
$\Lambda$-manifold (see Fig.~\ref{fig:trap-schematic}) to be pumped
back into the ground state by the 984 nm laser. By keeping the time
between measurement sufficiently long, the systematic effect
introduced during measurement was held below 10\%.

\section{T\lowercase{h}$^{3+}$ excitation effects on reaction rates}
\label{sec:optical-effects}

\begin{figure}
  \centering
  \includegraphics[angle=-90]{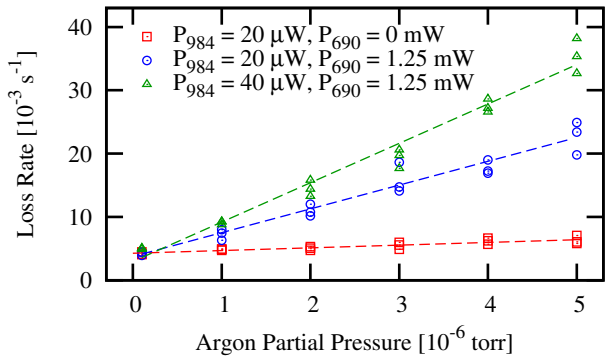}
  \caption{(Color online) Effect of atomic excitation on $\Thppp$ loss
    rate in the presence of argon.  The data that was taken without
    690 nm light represents loss from the $\Thppp$ $5\text{F}_{5/2}$
    ground state.  Increasing the power of the 984 nm laser increases
    the reaction rate by increasing the number of ions in the
    $6\text{D}_{5/2}$ excited state, from which reaction is more
    likely.}
  \label{fig:ArLightLossRates}
\end{figure}

\begin{figure}
  \centering
  \includegraphics[angle=-90]{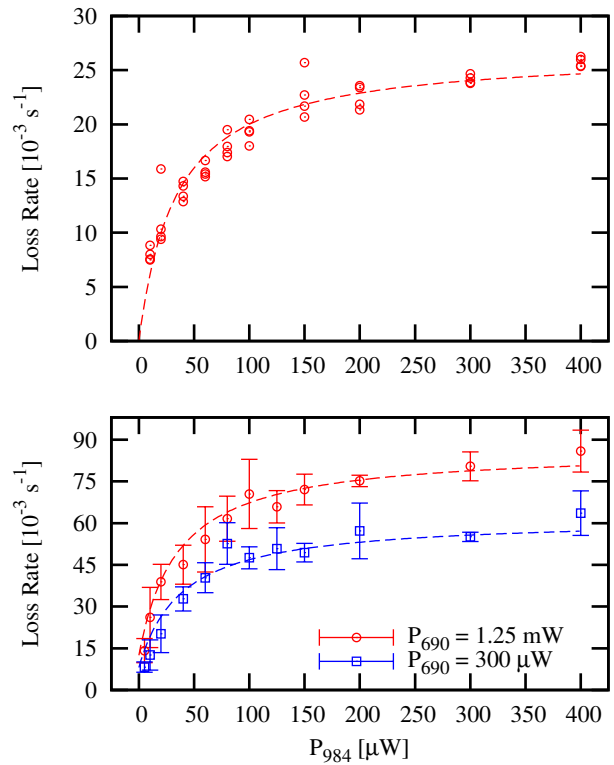}
  \caption{(Color online) Saturation of $\Thppp$ reaction rate in the
    presence of (top) $3 \times 10^{-6}$ torr argon and (bottom) $3
    \times 10^{-6}$ torr nitrogen. In each case, $1 \times 10^{-5}$
    torr helium is also present.}
  \label{fig:Saturation}
\end{figure}

State dependent effects on electron capture in collisions between
neutral atoms and multiply charged ions have been studied extensively
(see, e.g., \cite{Janev1985,Flechard2001,Knoop2005,Bodewits2007}). The
effect of Th$^{3+}$ optical excitation on its reaction rate with argon
can be seen clearly in Fig.~\ref{fig:ArLightLossRates}. In one set of
measurements, the 690 nm laser was shuttered between fluorescence
measurements. This is the same data set shown in
Fig.~\ref{fig:NeonArgon}. In the other two sets of measurements, the
690 nm laser was left on continuously. Increasing the power of the 984
nm laser increases the population of the 6D$_{5/2}$ excited state, and
hence the reaction rate, until the excited state is saturated. The
corresponding saturation of the argon and nitrogen reaction rates are
shown in Fig.~\ref{fig:Saturation}.

If the 5F$_{7/2}$--6D$_{5/2}$ manifold is approximated as a two-level
system, the excited state population obeys the relation
\begin{equation}
  \label{eq:state-pop}
  \sigma \propto \frac{I/I_{\mathrm{sat,eff}}}{1 + I/I_{\mathrm{sat,eff}}}.
\end{equation}
where $I_{\mathrm{sat,eff}}$ is the effective saturation intensity.
The ratio of the effective saturation intensity to the natural
saturation intensity, $I_{\mathrm{sat}} = 2\pi^2 hc\Gamma/3\lambda^3$,
is approximately equal to the ratio of the Doppler-broadened
linewidth, $\Gamma_D$, to the natural linewidth, $\Gamma$.  Therefore,
\begin{equation}
  \label{eq:I-sat-eff}
  I_{\mathrm{sat,eff}} \approx \frac{2\pi^2 hc}{3\lambda^3} \Gamma_D.
\end{equation}
By fitting the data in Fig.~\ref{fig:Saturation} to a function in the
form of Eq.~(\ref{eq:state-pop}), we find an effective saturation
intensity of $I_{\mathrm{sat,eff}} \sim 40$ mW/cm$^2$.  According to
Eq.~(\ref{eq:I-sat-eff}), this corresponds to a Doppler-broadened
linewidth of $\Gamma_D \sim 300$ MHz, which is consistent with an
independent measurement of the transition linewidth.

The reaction rate from the 5F$_{7/2}$ electronic state was found to be
equal to within experimental error to the reaction rate from the
5F$_{5/2}$ state for both argon and nitrogen.  The 5F$_{7/2}$ state
was isolated for measurement by shuttering the 984 nm laser between
measurements, while continuously applying the 690 nm laser.  In this
way, the long-lived 5F$_{7/2}$ state was continuosly repopulated.  To
further demonstrate that optical excitation to the 6D$_{5/2}$ state
was responsible for the increased reaction rates, we verified that
detuning the lasers far from resonance had the same effect as
shuttering them.

By optically exciting with the 1087 nm laser, we found that reactions
with nitrogen were also faster from the 6D$_{3/2}$ state than from the
5F$_{5/2}$ state.  No quantitative comparisons were made between the
reaction rates from the 6D$_{3/2}$ state and the 6D$_{5/2}$ state.
The reaction rate between Th$^{3+}$ and argon with 1087 nm optical
excitation was not measured.

\section{Gases with high reaction rates}
\label{sec:high-rate-gases}

\begin{figure}
  \centering
  \includegraphics[angle=-90]{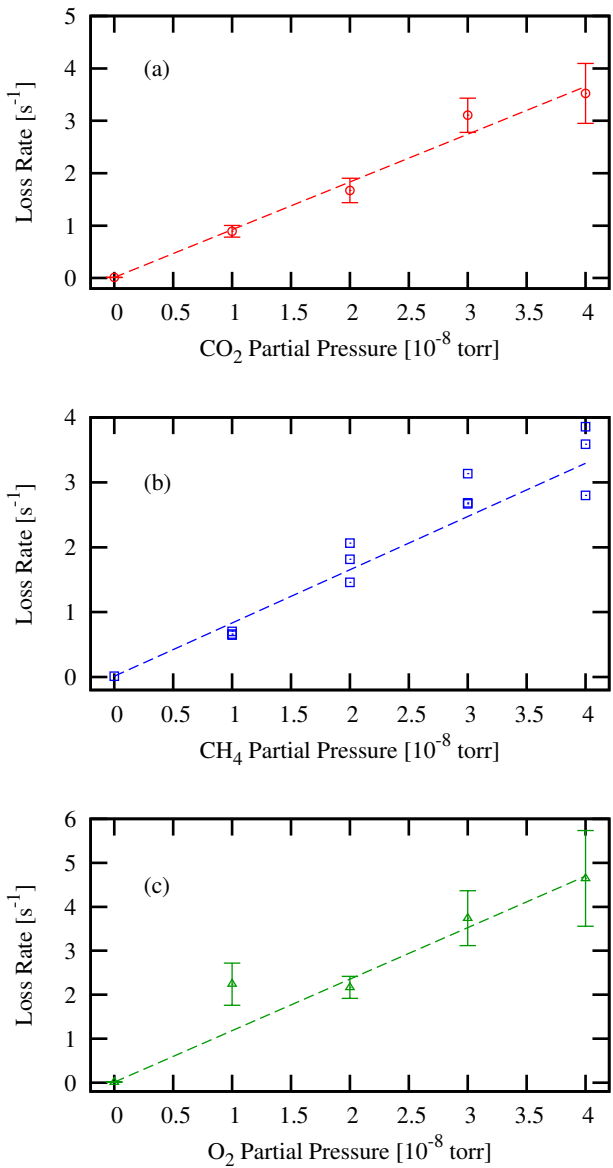}
  \caption{(Color online) $\Thppp$ loss rate in the presence of (a)
    carbon dioxide, (b) methane, and (c) oxygen. Measurements were
    made with $10^{-4}\,\text{torr}$ of helium buffer gas.  The data
    point at zero pressure is the loss rate when only helium at
    $10^{-4}\,\text{torr}$ is present. Reaction rate coefficients for
    each case are listed in Table~\ref{tab:rates}.}
  \label{fig:CO2MethaneOxygen}
\end{figure}

Loss rates of Th$^{3+}$ in the presence of carbon dioxide, methane,
and oxygen are shown in Fig.~\ref{fig:CO2MethaneOxygen}.  All of these
gases react relatively quickly with Th$^{3+}$ at rates comparable to
the Langevin rate.  Even at partial pressures below $10^{-8}$ torr,
reactions occurring in the space between the ablation target and the
trapping region significantly reduced the number of Th$^{3+}$ ions
loaded.  To ensure sufficient loading and an adequate SNR for
fluorescence measurement, $10^{-4}$ torr of helium was added to the
system for each measurement.  The loss rates measured here were in all
cases significantly higher than the loss rate with only $10^{-4}$ torr
of helium present.

The products of reaction between Th$^{3+}$ and carbon dioxide,
methane, and oxygen were identified using mass selective CEM
detection.  For these measurements, $2 \times 10^{-8}$ torr of the
test gas and $10^{-4}$ torr of helium buffer gas was introduced into
the system.  The ion trap was loaded a number of times.  Each time the
trap was loaded, we waited a short period ($\sim 1$ s) for reactions
to occur, performed a mass selection, and checked for the presence of
ions.  Since the rf power supply for our trap could not provide high
enough voltage to perform rigorous mass selection on singly ionized
molecules, we were able to identify only doubly ionized molecules.

When either carbon dioxide or oxygen was present in the system,
Th$^{2+}$ and ThO$^{2+}$ were found in the trap.  Although thorium
dioxide is chemically stable, no ThO$_2^{2+}$ was detected with either
gas.  In a previous work with singly-ionized thorium, Johnsen et
al. \cite{Johnsen1974} found that oxidation occurs via the sequential
reactions
\begin{eqnarray}
  \label{eq:Th-plus-O2}
  \mathrm{Th}^+ + \mathrm{O}_2 &\rightarrow& \mathrm{ThO}^+ + \mathrm{O} \\
  \label{eq:ThO-plus-O2}
  \mathrm{ThO}^+ + \mathrm{O}_2 &\rightarrow& \mathrm{ThO}_2^+ + \mathrm{O}
\end{eqnarray}
They found that the first reaction proceeded quickly, with a rate
coefficient of $6 \times 10^{-10}$ cm$^3$s$^{-1}$, while the rate of
the second reaction was more than an order of magnitude lower, with a
coefficient of $2 \times 10^{-11}$ cm$^3$s$^{-1}$.

Reactions between Th$^{3+}$ and methane, CH$_4$, resulted in Th$^{2+}$
and ThCH$_2^{2+}$.  Special care was taken to properly identify
ThCH$_2^{2+}$.  Andrews and Cho \cite{Andrews2005} were able to create
ThCH$_4$ by ablating thorium in a methane environment.  However,
Mar\c{c}alo et al. \cite{Marcalo1996} found that the only reaction
channel between singly-ionized thorium and methane is
\begin{equation}
  \label{eq:Th-plus-methane}
  \mathrm{Th}^+ + \mathrm{CH}_4 \rightarrow \mathrm{ThCH}_2^+ + \mathrm{H}_2.
\end{equation}

\begin{figure}
  \centering
  \includegraphics[angle=-90]{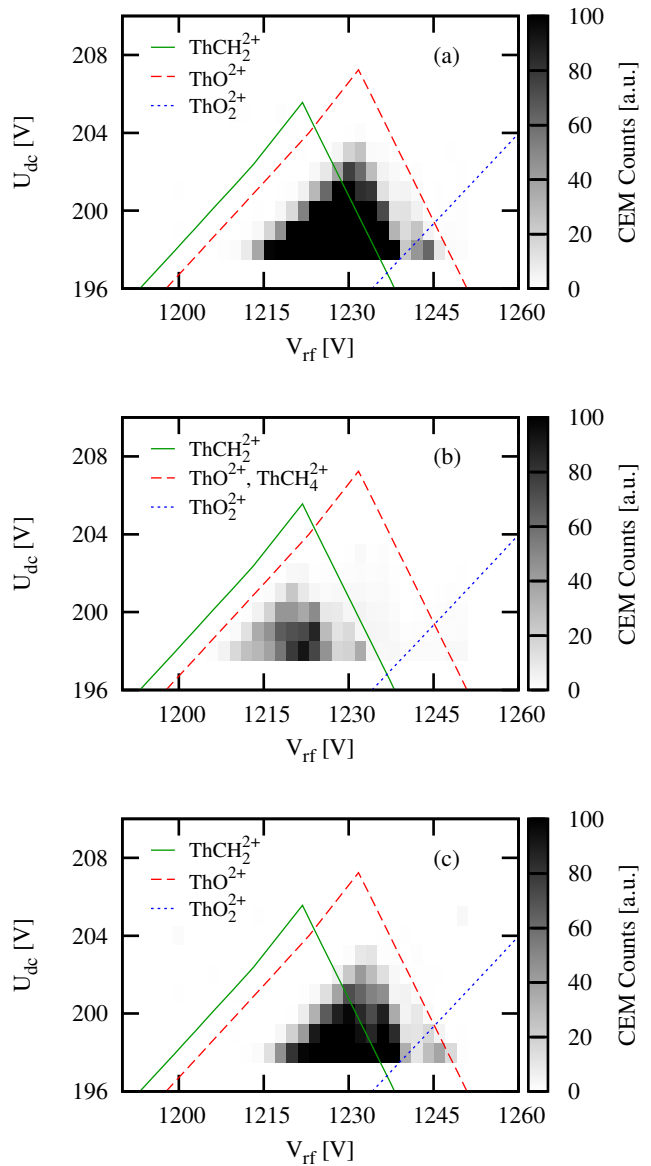}
  \caption{(Color online) Doubly-ionized thorium molecules.  Ions are
    identified by comparing the voltage space in which they are stable
    to the theoretical stability region.  (a) The reaction between
    $\Thppp$ and carbon dioxide produces $\text{ThO}^{2+}$. (b) The
    reaction between $\Thppp$ and methane produces $\Thpp$, which in
    turn reacts with methane to produce $\text{ThCH}_{2}^{2+}$. (c)
    The reaction between $\Thppp$ and oxygen produces $\Thpp$, which
    in turn reacts with oxygen to produce $\text{ThO}^{2+}$.}
  \label{fig:MassResolution}
\end{figure}

Only 1 amu/$e$ separates the mass-to-charge ratio of ThCH$_2^{2+}$
from ThCH$_4^{2+}$ and ThO$^{2+}$.  By carefully mapping the voltage
space in which ThO$^{2+}$ and ThCH$_2^{2+}$ were stable, we were able
to achieve the resolution necessary to correctly determine the product
(see Fig.~\ref{fig:MassResolution}).

\begin{figure}
  \centering
  \includegraphics[angle=-90]{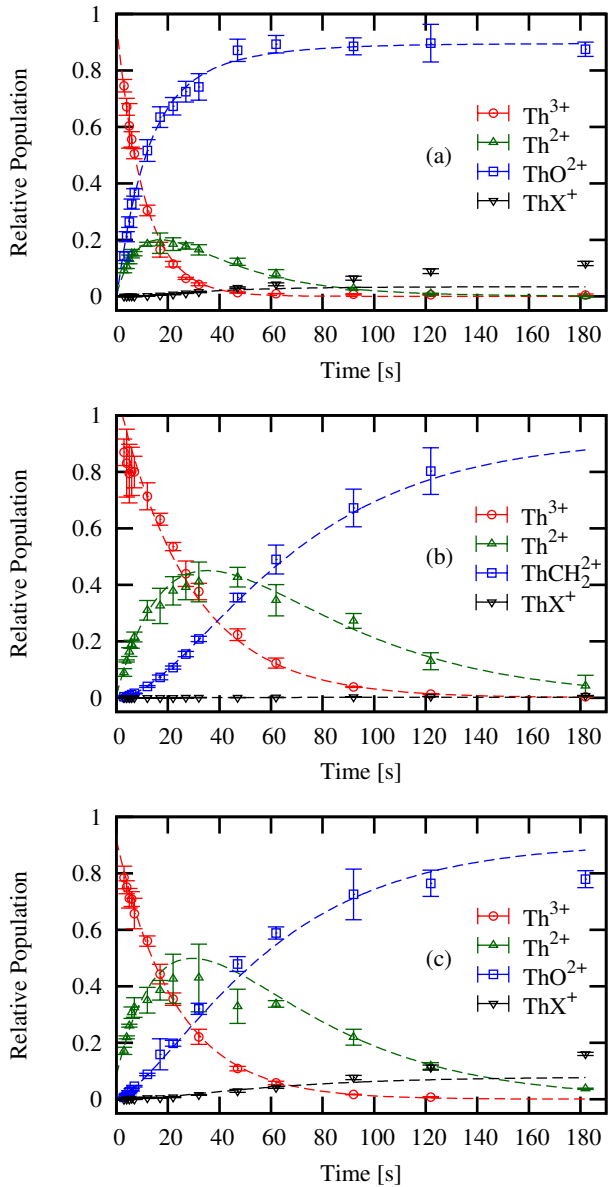}
  \caption{(Color online) $^{232}\Thppp$ in the presence of (a) carbon
    dioxide, (b) methane, and (c) oxygen. The data is scaled according
    to the values given in Table~\ref{tab:cem-scaling}. The data sets
    are fit to the system of equations given in
    Eqs.~(\ref{eq:fitth3})--(\ref{eq:fitthx1}).}
  \label{fig:Evolution}
\end{figure}

Once the reaction products were identified, we were able to measure
how Th$^{3+}$ evolved over time in the presence of each of these
gases.  The results are shown in Fig.~\ref{fig:Evolution} for carbon
dioxide, methane, and oxygen.  Each data point in these graphs
represents the average and standard deviation of six measurements.
For each measurement, the trap was loaded, and Th$^{3+}$ was mass
selected.  A fluorescence measurement would be taken immediately after
the mass selection.  After the specified wait time, the ion of
interest would be mass selected and the contents of the trap delivered
to the CEM.  The resulting CEM signal was normalized according to the
initial fluorescence measurement.  By normalizing the signal in this
way, the impact of loading fluctuations on the data was minimized.

As was mentioned previously, the rf power supply for our trap could
not provide high enough voltage to perform rigorous mass selection on
singly ionized molecules.  However, by ramping the rf to its maximum
voltage, we could isolate ions with mass-to-charge ratios greater than
156 amu/$e$.  The only ions in the trap above that mass-to-charge
ratio would be singly-ionized thorium molecules.  These are
represented as ThX$^+$ in Fig.~\ref{fig:Evolution}.

Direct comparison of the CEM signals of the various ions is of limited
value since the gain of a CEM can vary from ion to ion.  Ions with
higher charge states experience a greater acceleration in the field
created by the high negative bias at the entrance of the dynode.
These ions strike the dynode surface with greater energy, resulting in
a higher initial emission of electrons.  The CEM gain also depends on
mass, as lighter ions tend to cause a larger response than more
massive ones.  The chemical nature of the ion can also play a role.

We verified several times that the Th$^{3+}$ CEM signal was linear
with measurements of total fluorescence over the dynamic range of both
instruments.  However, the specific slope of the relationship at any
given time was highly sensitive to the background light level around
the system.  Curtains were employed to stabilize and minimize
background light; however, small shifts in the curtains incurred while
the system was prepared for data collection still had a noticeable
effect.

In order to compare the relative populations of the various ions in
the trap over time, we adopted a simple method for scaling the CEM
data.  The method is based on two assumptions.  We assume first that
ions are not lost from the trap over the period of investigation.
This is reasonable given that the trap depth is on the order of 100 eV
for the ions in question, while the exothermicity of the reactions is
only on the order of a few eV.  We further assume that the only ions
present in the trap are those identified in our earlier search for the
reaction products.  Since our method of searching for reaction
products focused only on products resulting from fast reactions, this
assumption could lead to error when considering longer time scales.
However, here we focus our attention on the rapid chemical kinetics
that occur on short time scales.

To scale the data we consider the sum
\begin{equation}
  \label{eq:trap-pop}
  N(t) = \sum_i \alpha_i n_i(t),
\end{equation}
where $N$ is the total trap population, and $n_i$ and $\alpha_i$ are
the fluorescence normalized CEM signal and the CEM scaling factor for
ion $i$, respectively.  The sum is over all ion species present in the
trap.  The scaling factors are determined via a least-squares
algorithm that attempts to make $N(t) = 1$ for all $t$.  The data
shown in Fig.~\ref{fig:Evolution} was scaled using this method.  The
scaling factors are given in Table~\ref{tab:cem-scaling}.

\begin{table}
  \centering
  \begin{ruledtabular}
    \begin{tabular}{c|c|c|c}
      \textbf{Species} & \textbf{Carbon Dioxide} & \textbf{Methane} &
      \textbf{Oxygen} \\ \hline
      $\Thppp$ & $9.0$ & $4.6$ & $3.6$ \\
      $\Thpp$ & $23.4$ & $26.2$ & $14.7$ \\
      ThO$^{2+}$ & $36.8$ & -- & $12.7$ \\
      ThCH$_{2}^{2+}$ & -- & $53.9$ & -- \\
      ThX$^+$ & $20.5$ & $30.0$ & $39.3$
    \end{tabular}
  \end{ruledtabular}
  \caption{CEM scaling factors for $\Thppp$ reaction data.}
  \label{tab:cem-scaling}
\end{table}

It is clear from Fig.~\ref{fig:Evolution} that the dynamics
responsible for the appearance of ThO$^{2+}$ in the presence of carbon
dioxide are different from the dynamics that cause ThO$^{2+}$ to arise
in oxygen and that bring about ThCH$_2^{2+}$ in the presence of
methane.  Based on the identified products, the possible reactions
between Th$^{3+}$ and these gases can be written generally as
\begin{eqnarray}
    \Thppp + \text{CO}_{2} &{}& \left\{
        \begin{array}{cl}
            \xrightarrow{\alpha_{1}}& \Thpp + \text{CO}_{2}^{+} \\
            \xrightarrow{\alpha_{2}}& \text{ThO}^{2+} + \text{CO}^{+}
        \end{array} \right. \\
    \Thppp + \text{CH}_{4} &{}& \left\{
        \begin{array}{cl}
            \xrightarrow{\beta_{1}}& \Thpp + \text{CH}_{4}^{+} \\
            \xrightarrow{\beta_{2}}& \text{ThCH}_{2}^{2+} + \text{H}_{2}^{+}
        \end{array} \right.\\
    \Thppp + \text{O}_{2} &{}& \left\{
        \begin{array}{cl}
            \xrightarrow{\gamma_{1}}& \Thpp + \text{O}_{2}^{+} \\
            \xrightarrow{\gamma_{2}}& \text{ThO}^{2+} + \text{O}^{+}
        \end{array} \right.
\end{eqnarray}
Here, $\alpha_i$, $\beta_i$, and $\gamma_i$ represent the relative
probabilities of the reaction channels.  The possible subsequent
reactions with $\Thpp$ can be summarized as
\begin{eqnarray}
    \Thpp + \text{CO}_{2} &{}& \left\{
        \begin{array}{cl}
            \xrightarrow{\alpha_{3}}& \text{Th}X^{+} + \cdots \\
            \xrightarrow{\alpha_{4}}& \text{ThO}^{2+} + \text{CO}
        \end{array} \right. \\
    \Thpp + \text{CH}_{4} &{}& \left\{
        \begin{array}{cl}
            \xrightarrow{\beta_{3}}& \text{Th}X^{+} + \cdots \\
            \xrightarrow{\beta_{4}}& \text{ThCH}_{2}^{2+} + \text{H}_{2}
        \end{array} \right. \\
    \Thpp + \text{O}_{2} &{}& \left\{
        \begin{array}{cl}
            \xrightarrow{\gamma_{3}}& \text{Th}X^{+} + \cdots \\
            \xrightarrow{\gamma_{4}}& \text{ThO}^{2+} + \text{O}
        \end{array} \right.
\end{eqnarray}
No decay in the quantity of ThO$^{2+}$ or ThCH$_2^{2+}$ is seen over
the time scales investigated, so these can be treated as stable final
products.  Thus, from the above reaction equations, we can write a
system of differential equations describing the chemical kinetics in
each gas.  For example, in carbon dioxide,
\begin{eqnarray}
    \ddt \func{n}{\Thppp}
        &=& -k_{1} \func{n}{\Thppp} \label{eq:fitth3}\\
    \ddt \func{n}{\Thpp}
        &=& -k_{2} \func{n}{\Thpp} + \alpha_{1} k_{1} \func{n}{\Thppp} \\
    \ddt \func{n}{\text{ThO}^{2+}}
        &=& \alpha_{2} k_{1} \func{n}{\Thppp} + \alpha_{4} k_{2} \func{n}{\Thpp} \\
    \ddt \func{n}{\text{Th}X^{+}}
        &=& \alpha_{3} k_{2} \func{n}{\Thpp} \label{eq:fitthx1}
\end{eqnarray}
Similar systems of equations can be written to describe the reaction
dynamics in methane and oxygen.

To estimate the relative probabilities of the various reaction
channels, the data of Fig.~\ref{fig:Evolution} was numerically fit to
dynamic equations of the form shown above.  Fits were performed in the
order of the above equations.  Once a value was obtained for a fitting
parameter, that value was enforced on subsequent fits.  The branching
ratios that were determined in this fashion are given in
Table~\ref{tab:branching-ratios}.

\begin{table}
  \centering
  \begin{ruledtabular}
    \begin{tabular}{c|c|c|c}
      \textbf{Branching Ratio} & \textbf{Carbon Dioxide} &
      \textbf{Methane} & \textbf{Oxygen} \\ \hline
      $\left[ \alpha, \beta, \gamma \right]_{1}$ &
      $0.36$ & $0.89$ & $1.0$ \\
      $\left[ \alpha, \beta, \gamma \right]_{2}$ &
      $0.78$ & $0.0$ & $0.09$ \\
      $\left[ \alpha, \beta, \gamma \right]_{3}$ &
      $0.10$ & $0.0$ & $0.08$ \\
      $\left[ \alpha, \beta, \gamma \right]_{4}$ &
      $0.44$ & $0.94$ & $0.83$
    \end{tabular}
  \end{ruledtabular}
  \caption{Branching ratios for thorium reactions.}
  \label{tab:branching-ratios}
\end{table}

The values given in Table~\ref{tab:branching-ratios} can only be
considered rough estimates of the actual branching ratios because of
the uncertainties inherent in the CEM scaling procedure.  They are
given only to illustrate the general reaction dynamics.  While
reactions of $\Thppp$ with methane and oxygen proceed predominantly
via charge exchange, simultaneous charge exchange and chemical
reaction is the dominant branch in reactions between $\Thppp$ and
carbon dioxide.  Since methane has the highest ionization energy and
lowest polarizability among hydrocarbons \cite{Lide2007}, the charge
exchange reaction rates between other hydrocarbons and $\Thppp$ are
likely as high as those of methane.


It is worth considering whether impurities of these fast reacting
gases could explain the $\Thppp$ losses in the presence of argon,
nitrogen, and hydrogen. CO$_2$, O$_2$, and CH$_4$ each react strongly
with $\Thppp$ at rates very close to the calculated Langevin rates. In
order for these molecules to account for the observed loss rates in
the other gases, they would have to be present as impurities at the
$10^{-4}$ level reacting at the Langevin rate, considerably higher
than the levels specified by the gas supplier (see
Table~\ref{tab:gas-purity}). Furthermore, none of these molecules
exhibited different reaction rates due to optical excitation of
$\Thppp$, so their presence as impurities could not explain the
reaction rates observed for argon and nitrogen. For these reasons, it
is unlikely that impurities of the type measured to react at the
Langevin rate are responsible for the losses observed for the slowly
reacting gases.

\begin{table}
  \centering
  \begin{ruledtabular}
    \begin{tabular}{c|c|c|c|c}
      \textbf{Gas} & \textbf{Minimum Purity} & $\mathbf{CO_2}$ &
      $\mathbf{O_2}$ & \textbf{Hydrocarbons} \\ \hline
      Argon & 99.9997\% & $< 0.5$ & $< 0.2$ & $< 0.2$
      \\ 
      Hydrogen & 99.999\% & $< 0.5$ & $< 1$ & $< 0.5$ \\
      Neon & 99.999\% & $< 0.5$ & $< 0.5$ & $< 0.5$
    \end{tabular}
    \caption{Purity specifications for research grade argon, ultrahigh
      purity hydrogen, and research grade neon used in reaction
      experiments. Values listed are in ppm. All gases were purchased
      from Airgas, Inc. and specifications are listed in the Airgas
      2009 Product Catalog \cite{Airgas2009}.}
  \end{ruledtabular}
  \label{tab:gas-purity}
\end{table}

\section{Conclusion}
\label{sec:conclusion}

We have determined the effect of various gases on the trap lifetime of
$\Thppp$.  Reactions involving carbon dioxide, oxygen, and methane
proceed at a rate near the classical Langevin limit.  Although charge
exchange reactions with them are exothermic, the reaction rate
coefficients for nitrogen, hydrogen, and argon are orders of magnitude
less than the Langevin rate.  The reaction rate coefficient between
$\Thppp$ and helium provides a measure of the impurities in the buffer
gas.  The measurement of the coefficient here provides a standard
against which the vacuum and buffer gas quality of future systems can
be measured.

\acknowledgments We gratefully acknowledge Adam Steele, Corey
Campbell, Alex Radnaev, and Alex Kuzmich for their assistance with
this work, and we thank Ken Brown for valuable discussions. This work
was supported by the Office of Naval Research (N000140911024) and the
National Science Foundation (PHYS-1002550).


\begin{thebibliography}{36}
\expandafter\ifx\csname natexlab\endcsname\relax\def\natexlab#1{#1}\fi
\expandafter\ifx\csname bibnamefont\endcsname\relax
  \def\bibnamefont#1{#1}\fi
\expandafter\ifx\csname bibfnamefont\endcsname\relax
  \def\bibfnamefont#1{#1}\fi
\expandafter\ifx\csname citenamefont\endcsname\relax
  \def\citenamefont#1{#1}\fi
\expandafter\ifx\csname url\endcsname\relax
  \def\url#1{\texttt{#1}}\fi
\expandafter\ifx\csname urlprefix\endcsname\relax\def\urlprefix{URL }\fi
\providecommand{\bibinfo}[2]{#2}
\providecommand{\eprint}[2][]{\url{#2}}

\bibitem[{\citenamefont{Beck et~al.}(2007)\citenamefont{Beck, Becker,
  Beiersdorfer, Brown, Moody, Wilhelmy, Porter, Kilbourne, and
  Kelley}}]{Beck2007}
\bibinfo{author}{\bibfnamefont{B.~R.} \bibnamefont{Beck}},
  \bibinfo{author}{\bibfnamefont{J.~A.} \bibnamefont{Becker}},
  \bibinfo{author}{\bibfnamefont{P.}~\bibnamefont{Beiersdorfer}},
  \bibinfo{author}{\bibfnamefont{G.~V.} \bibnamefont{Brown}},
  \bibinfo{author}{\bibfnamefont{K.~J.} \bibnamefont{Moody}},
  \bibinfo{author}{\bibfnamefont{J.~B.} \bibnamefont{Wilhelmy}},
  \bibinfo{author}{\bibfnamefont{F.~S.} \bibnamefont{Porter}},
  \bibinfo{author}{\bibfnamefont{C.~A.} \bibnamefont{Kilbourne}},
  \bibnamefont{and} \bibinfo{author}{\bibfnamefont{R.~L.}
  \bibnamefont{Kelley}}, \bibinfo{journal}{Phys. Rev. Lett.}
  \textbf{\bibinfo{volume}{98}}, \bibinfo{eid}{142501} (\bibinfo{year}{2007}).

\bibitem[{\citenamefont{Peik and Tamm}(2003)}]{Peik2003}
\bibinfo{author}{\bibfnamefont{E.}~\bibnamefont{Peik}} \bibnamefont{and}
  \bibinfo{author}{\bibfnamefont{C.}~\bibnamefont{Tamm}},
  \bibinfo{journal}{Europhys. Lett.} \textbf{\bibinfo{volume}{61}},
  \bibinfo{pages}{181} (\bibinfo{year}{2003}).

\bibitem[{\citenamefont{Flambaum}(2006)}]{Flambaum2006}
\bibinfo{author}{\bibfnamefont{V.~V.} \bibnamefont{Flambaum}},
  \bibinfo{journal}{Phys. Rev. Lett.} \textbf{\bibinfo{volume}{97}},
  \bibinfo{eid}{092502} (\bibinfo{year}{2006}).

\bibitem[{\citenamefont{Hayes and Friar}(2007)}]{Hayes2007}
\bibinfo{author}{\bibfnamefont{A.~C.} \bibnamefont{Hayes}} \bibnamefont{and}
  \bibinfo{author}{\bibfnamefont{J.~L.} \bibnamefont{Friar}},
  \bibinfo{journal}{Phys. Lett. B} \textbf{\bibinfo{volume}{650}},
  \bibinfo{pages}{229} (\bibinfo{year}{2007}).

\bibitem[{\citenamefont{He and Ren}(2007)}]{He2007}
\bibinfo{author}{\bibfnamefont{X.~T.} \bibnamefont{He}} \bibnamefont{and}
  \bibinfo{author}{\bibfnamefont{Z.~Z.} \bibnamefont{Ren}},
  \bibinfo{journal}{Journal of Physics G-Nuclear and Particle Physics}
  \textbf{\bibinfo{volume}{34}}, \bibinfo{pages}{1611} (\bibinfo{year}{2007}).

\bibitem[{\citenamefont{He and Ren}(2008)}]{He2008}
\bibinfo{author}{\bibfnamefont{X.~T.} \bibnamefont{He}} \bibnamefont{and}
  \bibinfo{author}{\bibfnamefont{Z.~Z.} \bibnamefont{Ren}},
  \bibinfo{journal}{Nucl. Phys. A} \textbf{\bibinfo{volume}{806}},
  \bibinfo{pages}{117} (\bibinfo{year}{2008}).

\bibitem[{\citenamefont{Berengut et~al.}(2009)\citenamefont{Berengut, Dzuba,
  Flambaum, and Porsev}}]{Berengut2009}
\bibinfo{author}{\bibfnamefont{J.~C.} \bibnamefont{Berengut}},
  \bibinfo{author}{\bibfnamefont{V.~A.} \bibnamefont{Dzuba}},
  \bibinfo{author}{\bibfnamefont{V.~V.} \bibnamefont{Flambaum}},
  \bibnamefont{and} \bibinfo{author}{\bibfnamefont{S.~G.}
  \bibnamefont{Porsev}}, \bibinfo{journal}{Phys. Rev. Lett.}
  \textbf{\bibinfo{volume}{102}}, \bibinfo{eid}{210801} (\bibinfo{year}{2009}).

\bibitem[{\citenamefont{Campbell et~al.}(2009)\citenamefont{Campbell, Steele,
  Churchill, DePalatis, Naylor, Matsukevich, Kuzmich, and
  Chapman}}]{Campbell2009}
\bibinfo{author}{\bibfnamefont{C.~J.} \bibnamefont{Campbell}},
  \bibinfo{author}{\bibfnamefont{A.~V.} \bibnamefont{Steele}},
  \bibinfo{author}{\bibfnamefont{L.~R.} \bibnamefont{Churchill}},
  \bibinfo{author}{\bibfnamefont{M.~V.} \bibnamefont{DePalatis}},
  \bibinfo{author}{\bibfnamefont{D.~E.} \bibnamefont{Naylor}},
  \bibinfo{author}{\bibfnamefont{D.~N.} \bibnamefont{Matsukevich}},
  \bibinfo{author}{\bibfnamefont{A.}~\bibnamefont{Kuzmich}}, \bibnamefont{and}
  \bibinfo{author}{\bibfnamefont{M.~S.} \bibnamefont{Chapman}},
  \bibinfo{journal}{Phys. Rev. Lett.} \textbf{\bibinfo{volume}{102}},
  \bibinfo{eid}{233004} (\bibinfo{year}{2009}).

\bibitem[{\citenamefont{Langevin}(1903)}]{Langevin1903}
\bibinfo{author}{\bibfnamefont{P.}~\bibnamefont{Langevin}},
  \bibinfo{journal}{Annales De Chimie Et De Physique}
  \textbf{\bibinfo{volume}{28}}, \bibinfo{pages}{433} (\bibinfo{year}{1903}).

\bibitem[{\citenamefont{Gioumousis and Stevenson}(1958)}]{Gioumousis1958}
\bibinfo{author}{\bibfnamefont{G.}~\bibnamefont{Gioumousis}} \bibnamefont{and}
  \bibinfo{author}{\bibfnamefont{D.~P.} \bibnamefont{Stevenson}},
  \bibinfo{journal}{J. Chem. Phys.} \textbf{\bibinfo{volume}{29}},
  \bibinfo{pages}{294} (\bibinfo{year}{1958}).

\bibitem[{\citenamefont{Troe}(1985)}]{Troe1985}
\bibinfo{author}{\bibfnamefont{J.}~\bibnamefont{Troe}}, \bibinfo{journal}{Chem.
  Phys. Lett.} \textbf{\bibinfo{volume}{122}}, \bibinfo{pages}{425}
  (\bibinfo{year}{1985}).

\bibitem[{\citenamefont{Schissler and Stevenson}(1956)}]{Schissler1956}
\bibinfo{author}{\bibfnamefont{D.~O.} \bibnamefont{Schissler}}
  \bibnamefont{and} \bibinfo{author}{\bibfnamefont{D.~P.}
  \bibnamefont{Stevenson}}, \bibinfo{journal}{J. Chem. Phys.}
  \textbf{\bibinfo{volume}{24}}, \bibinfo{pages}{926} (\bibinfo{year}{1956}).

\bibitem[{\citenamefont{Field et~al.}(1957)\citenamefont{Field, Franklin, and
  Lampe}}]{Field1957}
\bibinfo{author}{\bibfnamefont{F.~H.} \bibnamefont{Field}},
  \bibinfo{author}{\bibfnamefont{J.~L.} \bibnamefont{Franklin}},
  \bibnamefont{and} \bibinfo{author}{\bibfnamefont{F.~W.} \bibnamefont{Lampe}},
  \bibinfo{journal}{J. Am. Chem. Soc.} \textbf{\bibinfo{volume}{79}},
  \bibinfo{pages}{2419} (\bibinfo{year}{1957}).

\bibitem[{\citenamefont{Boelrijk and Hamill}(1962)}]{Boelrijk1962}
\bibinfo{author}{\bibfnamefont{N.}~\bibnamefont{Boelrijk}} \bibnamefont{and}
  \bibinfo{author}{\bibfnamefont{W.~H.} \bibnamefont{Hamill}},
  \bibinfo{journal}{J. Am. Chem. Soc.} \textbf{\bibinfo{volume}{84}},
  \bibinfo{pages}{730} (\bibinfo{year}{1962}).

\bibitem[{\citenamefont{Mahan}(1971)}]{Mahan1971}
\bibinfo{author}{\bibfnamefont{B.~H.} \bibnamefont{Mahan}},
  \bibinfo{journal}{J. Chem. Phys.} \textbf{\bibinfo{volume}{55}},
  \bibinfo{pages}{1436} (\bibinfo{year}{1971}).

\bibitem[{\citenamefont{Johnsen and Biondi}(1978)}]{Johnsen1978}
\bibinfo{author}{\bibfnamefont{R.}~\bibnamefont{Johnsen}} \bibnamefont{and}
  \bibinfo{author}{\bibfnamefont{M.~A.} \bibnamefont{Biondi}},
  \bibinfo{journal}{Phys. Rev. A} \textbf{\bibinfo{volume}{18}},
  \bibinfo{pages}{996} (\bibinfo{year}{1978}).

\bibitem[{\citenamefont{Johnsen and Biondi}(1979)}]{Johnsen1979}
\bibinfo{author}{\bibfnamefont{R.}~\bibnamefont{Johnsen}} \bibnamefont{and}
  \bibinfo{author}{\bibfnamefont{M.~A.} \bibnamefont{Biondi}},
  \bibinfo{journal}{Phys. Rev. A} \textbf{\bibinfo{volume}{20}},
  \bibinfo{pages}{87} (\bibinfo{year}{1979}).

\bibitem[{\citenamefont{Johnsen and Biondi}(1980)}]{Johnsen1980}
\bibinfo{author}{\bibfnamefont{R.}~\bibnamefont{Johnsen}} \bibnamefont{and}
  \bibinfo{author}{\bibfnamefont{M.~A.} \bibnamefont{Biondi}},
  \bibinfo{journal}{J. Chem. Phys.} \textbf{\bibinfo{volume}{73}},
  \bibinfo{pages}{190} (\bibinfo{year}{1980}).

\bibitem[{\citenamefont{Holzscheiter and
  Church}(1981{\natexlab{a}})}]{Holzscheiter1981}
\bibinfo{author}{\bibfnamefont{H.~M.} \bibnamefont{Holzscheiter}}
  \bibnamefont{and} \bibinfo{author}{\bibfnamefont{D.~A.}
  \bibnamefont{Church}}, \bibinfo{journal}{J. Chem. Phys.}
  \textbf{\bibinfo{volume}{74}}, \bibinfo{pages}{2313}
  (\bibinfo{year}{1981}{\natexlab{a}}).

\bibitem[{\citenamefont{Church and Holzscheiter}(1989)}]{Church1989}
\bibinfo{author}{\bibfnamefont{D.~A.} \bibnamefont{Church}} \bibnamefont{and}
  \bibinfo{author}{\bibfnamefont{H.~M.} \bibnamefont{Holzscheiter}},
  \bibinfo{journal}{Phys. Rev. A} \textbf{\bibinfo{volume}{40}},
  \bibinfo{pages}{54} (\bibinfo{year}{1989}).

\bibitem[{\citenamefont{McDaniel}(1970)}]{McDaniel1970}
\bibinfo{author}{\bibfnamefont{E.~W.} \bibnamefont{McDaniel}},
  \emph{\bibinfo{title}{Ion-molecule reactions}}, Wiley-Interscience series in
  atomic and molecular collisional processes
  (\bibinfo{publisher}{Wiley-Interscience}, \bibinfo{address}{New York},
  \bibinfo{year}{1970}).

\bibitem[{\citenamefont{Johnsen et~al.}(1974)\citenamefont{Johnsen, Castell,
  and Biondi}}]{Johnsen1974}
\bibinfo{author}{\bibfnamefont{R.}~\bibnamefont{Johnsen}},
  \bibinfo{author}{\bibfnamefont{F.~R.} \bibnamefont{Castell}},
  \bibnamefont{and} \bibinfo{author}{\bibfnamefont{M.~A.}
  \bibnamefont{Biondi}}, \bibinfo{journal}{J. Chem. Phys.}
  \textbf{\bibinfo{volume}{61}}, \bibinfo{pages}{5404} (\bibinfo{year}{1974}).

\bibitem[{\citenamefont{Holzscheiter and
  Church}(1981{\natexlab{b}})}]{Holzscheiter1981a}
\bibinfo{author}{\bibfnamefont{H.~M.} \bibnamefont{Holzscheiter}}
  \bibnamefont{and} \bibinfo{author}{\bibfnamefont{D.~A.}
  \bibnamefont{Church}}, \bibinfo{journal}{Phys. Lett. A}
  \textbf{\bibinfo{volume}{86}}, \bibinfo{pages}{25}
  (\bibinfo{year}{1981}{\natexlab{b}}).

\bibitem[{\citenamefont{Chatham et~al.}(1983)\citenamefont{Chatham, Hils,
  Robertson, and Gallagher}}]{Chatham1983}
\bibinfo{author}{\bibfnamefont{H.}~\bibnamefont{Chatham}},
  \bibinfo{author}{\bibfnamefont{D.}~\bibnamefont{Hils}},
  \bibinfo{author}{\bibfnamefont{R.}~\bibnamefont{Robertson}},
  \bibnamefont{and} \bibinfo{author}{\bibfnamefont{A.~C.}
  \bibnamefont{Gallagher}}, \bibinfo{journal}{J. Chem. Phys.}
  \textbf{\bibinfo{volume}{79}}, \bibinfo{pages}{1301} (\bibinfo{year}{1983}).

\bibitem[{\citenamefont{Church}(1992)}]{Church1992}
\bibinfo{author}{\bibfnamefont{D.~A.} \bibnamefont{Church}},
  \bibinfo{journal}{J. Mod. Opt.} \textbf{\bibinfo{volume}{39}},
  \bibinfo{pages}{423} (\bibinfo{year}{1992}).

\bibitem[{\citenamefont{Andrews and Cho}(2005)}]{Andrews2005}
\bibinfo{author}{\bibfnamefont{L.}~\bibnamefont{Andrews}} \bibnamefont{and}
  \bibinfo{author}{\bibfnamefont{H.~G.} \bibnamefont{Cho}},
  \bibinfo{journal}{J. Phys. Chem. A} \textbf{\bibinfo{volume}{109}},
  \bibinfo{pages}{6796} (\bibinfo{year}{2005}).

\bibitem[{\citenamefont{Major and Dehmelt}(1968)}]{Major1968}
\bibinfo{author}{\bibfnamefont{F.~G.} \bibnamefont{Major}} \bibnamefont{and}
  \bibinfo{author}{\bibfnamefont{H.~G.} \bibnamefont{Dehmelt}},
  \bibinfo{journal}{Phys. Rev.} \textbf{\bibinfo{volume}{170}},
  \bibinfo{pages}{91} (\bibinfo{year}{1968}).

\bibitem[{\citenamefont{Wu and Brodbelt}(1993)}]{Wu1993}
\bibinfo{author}{\bibfnamefont{H.}~\bibnamefont{Wu}} \bibnamefont{and}
  \bibinfo{author}{\bibfnamefont{J.~S.} \bibnamefont{Brodbelt}},
  \bibinfo{journal}{Int. J. Mass Spectrom. Ion Processes}
  \textbf{\bibinfo{volume}{124}}, \bibinfo{pages}{175} (\bibinfo{year}{1993}).

\bibitem[{\citenamefont{DeVoe}(2009)}]{DeVoe2009}
\bibinfo{author}{\bibfnamefont{R.~G.} \bibnamefont{DeVoe}},
  \bibinfo{journal}{Physical Review Letters} \textbf{\bibinfo{volume}{102}},
  \bibinfo{eid}{063001} (\bibinfo{year}{2009}).

\bibitem[{\citenamefont{{Airgas, Inc.}}(2009)}]{Airgas2009}
\bibinfo{author}{\bibnamefont{{Airgas, Inc.}}}, \emph{\bibinfo{title}{Airgas
  2009 {P}roduct {C}atalog}} (\bibinfo{year}{2009}).

\bibitem[{\citenamefont{Janev and Winter}(1985)}]{Janev1985}
\bibinfo{author}{\bibfnamefont{R.~K.} \bibnamefont{Janev}} \bibnamefont{and}
  \bibinfo{author}{\bibfnamefont{H.}~\bibnamefont{Winter}},
  \bibinfo{journal}{Physics Reports} \textbf{\bibinfo{volume}{117}},
  \bibinfo{pages}{265} (\bibinfo{year}{1985}).

\bibitem[{\citenamefont{Fl\'{e}chard et~al.}(2001)\citenamefont{Fl\'{e}chard,
  Harel, Jouin, Pons, Adoui, Fr\'{e}mont, Cassimi, and
  Hennecart}}]{Flechard2001}
\bibinfo{author}{\bibfnamefont{X.}~\bibnamefont{Fl\'{e}chard}},
  \bibinfo{author}{\bibfnamefont{C.}~\bibnamefont{Harel}},
  \bibinfo{author}{\bibfnamefont{H.}~\bibnamefont{Jouin}},
  \bibinfo{author}{\bibfnamefont{B.}~\bibnamefont{Pons}},
  \bibinfo{author}{\bibfnamefont{L.}~\bibnamefont{Adoui}},
  \bibinfo{author}{\bibfnamefont{F.}~\bibnamefont{Fr\'{e}mont}},
  \bibinfo{author}{\bibfnamefont{A.}~\bibnamefont{Cassimi}}, \bibnamefont{and}
  \bibinfo{author}{\bibfnamefont{D.}~\bibnamefont{Hennecart}},
  \bibinfo{journal}{J. Phys. B: At., Mol. Opt. Phys.}
  \textbf{\bibinfo{volume}{34}}, \bibinfo{pages}{2759} (\bibinfo{year}{2001}).

\bibitem[{\citenamefont{Knoop et~al.}(2005)\citenamefont{Knoop, Keim, Lüdde,
  Kirchner, Morgenstern, and Hoekstra}}]{Knoop2005}
\bibinfo{author}{\bibfnamefont{S.}~\bibnamefont{Knoop}},
  \bibinfo{author}{\bibfnamefont{M.}~\bibnamefont{Keim}},
  \bibinfo{author}{\bibfnamefont{H.~J.} \bibnamefont{Lüdde}},
  \bibinfo{author}{\bibfnamefont{T.}~\bibnamefont{Kirchner}},
  \bibinfo{author}{\bibfnamefont{R.}~\bibnamefont{Morgenstern}},
  \bibnamefont{and} \bibinfo{author}{\bibfnamefont{R.}~\bibnamefont{Hoekstra}},
  \bibinfo{journal}{J. Phys. B: At., Mol. Opt. Phys.}
  \textbf{\bibinfo{volume}{38}}, \bibinfo{pages}{3163} (\bibinfo{year}{2005}).

\bibitem[{\citenamefont{Bodewits and Hoekstra}(2007)}]{Bodewits2007}
\bibinfo{author}{\bibfnamefont{D.}~\bibnamefont{Bodewits}} \bibnamefont{and}
  \bibinfo{author}{\bibfnamefont{R.}~\bibnamefont{Hoekstra}},
  \bibinfo{journal}{Phys. Rev. A} \textbf{\bibinfo{volume}{76}},
  \bibinfo{pages}{032703} (\bibinfo{year}{2007}).

\bibitem[{\citenamefont{Mar\c{c}alo et~al.}(1996)\citenamefont{Mar\c{c}alo,
  Leal, and de~Matos}}]{Marcalo1996}
\bibinfo{author}{\bibfnamefont{J.}~\bibnamefont{Mar\c{c}alo}},
  \bibinfo{author}{\bibfnamefont{J.~P.} \bibnamefont{Leal}}, \bibnamefont{and}
  \bibinfo{author}{\bibfnamefont{A.~P.} \bibnamefont{de~Matos}},
  \bibinfo{journal}{International Journal of Mass Spectrometry and Ion
  Processes} \textbf{\bibinfo{volume}{157-158}}, \bibinfo{pages}{265 }
  (\bibinfo{year}{1996}), ISSN \bibinfo{issn}{0168-1176}.

\bibitem[{\citenamefont{Lide}(2007)}]{Lide2007}
\bibinfo{author}{\bibfnamefont{D.~R.} \bibnamefont{Lide}},
  \emph{\bibinfo{title}{CRC Handbook of Chemistry and Physics}}
  (\bibinfo{publisher}{CRC}, \bibinfo{year}{2007}), \bibinfo{edition}{88th}
  ed., ISBN \bibinfo{isbn}{0849304881}.

\end{thebibliography}
\end{document}